\documentclass[usegraphicx,useAMS,usenatbib]{mn2e}

\newcommand{\hi}{H\,{\sc i}}
\newcommand{\kms}{km\,s$^{-1}$}

\title[Polar BAL Quasars]{Polar Broad Absorption Line Quasars: An Open Question}
\author[P. B. Hall, L. S. Chajet]{P. B. 
Hall$^{1}$\thanks{E-mail: phall@yorku.ca (PBH); lchajet@yorku.ca (LSC)},
L. S. Chajet$^1$\footnotemark[1]\\
$^1$Department of Physics and Astronomy, York University, 4700 Keele St., Toronto, Ontario M3J 1P3, Canada} 
\begin{document}


\pagerange{\pageref{firstpage}--\pageref{lastpage}} \pubyear{2002}

\maketitle

\label{firstpage}

\begin{abstract}
It has been argued that certain broad absorption line quasars are viewed
within 35$^\circ$ of the axis of a relativistic radio jet, based on
two-epoch radio flux density variability.
It is true if the surface brightness of a radio source is
observed to change by a sufficiently large amount, the inferred
brightness temperature will exceed $10^{12}$\,K and Doppler beaming 
in our direction must be invoked to avoid a Compton cooling catastrophe.
However, flux density changes cannot be linked to surface brightness changes
without knowledge of the size of the source.
If an optically thick source changes in projected area but not surface
brightness, its brightness temperature is constant and its flux variability
yields no constraint on its orientation.  Moreover, as pointed out by Rees,
spherical expansion of an emission source at relativistic speeds yields
an apparently superluminal increase in its projected area, which can explain
short-timescale flux density variability without requiring a relativistic
jet oriented near to our line of sight.
Therefore, two-epoch radio flux density variability by itself cannot 
unambiguously identify sources with jets directed toward us.
Only VLBI imaging can robustly determine 
the fraction of broad absorption line quasars which are polar.
\end{abstract}


\section{Introduction}

Broad absorption line (BAL) quasars are those quasars which show ultraviolet
absorption troughs thousands of km\,s$^{-1}$ in width and outflow velocity
\citep[e.g.,][]{allenbal}.
Models for BAL quasars have often relied on winds from accretion disks,
in which BAL quasars are more likely to be observed at low latitudes above
the disk than at high latitudes (e.g., \citealt{dkb95,mc95,pk04,fkcb10}).
It was therefore notable when \citet[][hereafter Z+06]{zhou+06} reported 
that six BAL quasars from the Sloan Digital Sky Survey \citep{yor00}
were ``polar'', constrained by radio flux density
variability to lie within 35$^\circ$ of a relativistic jet 
(presumably located at 90$^\circ$ latitude above the accretion disk).  
Shortly thereafter, \citet[][hereafter GP07]{gp07} 
presented nine additional candidate polar BAL quasars,
\citet{2008MNRAS.388.1853M} and \citet{2009PASJ...61.1389D}
noted two new candidates each, and \citet{2009ApJ...706..851R} 
suggested Mrk~231 as a candidate.

The approach of using radio flux density variability to constrain a quasar's
orientation is based upon the existence of a limiting brightness temperature 
$T_b$.  If $T_b > 10^{12}$\,K in a radio source, a Compton cooling
catastrophe occurs.  The timescale for the electrons producing the radio 
synchrotron emission to lose all their energy through inverse Compton 
scattering of ambient radio photons decreases to just a few days, and the
brightness temperature falls back below $10^{12}$\,K \citep{kpt69}.
(In fact, the upper limit brightness temperature may be closer to
$T_b\simeq 10^{11}$\,K; see \citealt{1994ApJ...426...51R}.)
To explain cases where $T_b > 10^{12} \rm ~K$ is observed,
Doppler boosting in our direction is invoked.
Relativistic motion of the radio-emitting plasma in our direction,
e.g. in a relativistic jet oriented close to our line of sight,
will boost the flux and brightness temperature we infer \citep{lb85}.
The brightness temperature is related to the surface brightness; therefore, a
sufficiently large change in the surface brightness of a radio source indicates
an emission component with a brightness temperature $T_b > 10^{12}$\,K
and Doppler boosting in our direction.

As knowledge of the incidence of BALs as a function of latitude would be a
useful constraint on models of BAL outflows, close analysis of the assumptions
leading to the designation of some BAL quasars as ``polar'' is worthwhile.
We present a brief such analysis in \S \ref{analysis}.  
First, in \S \ref{data} we correct an error in the literature that led to 
overestimates of the significance of variability in some ``polar'' BAL quasars.

\section{Data}\label{data}

The data used in Z+06, GP07, \citet{2008MNRAS.388.1853M},
\citet{2009PASJ...61.1389D} and herein comes from the 
FIRST \citep{bwh95,1997ApJ...475..479W} and NVSS \citep{con98} catalogs.
In Table 1 we report peak flux densities and their uncertainties 
from both surveys for all 21 polar BAL quasars reported to date.
For ease of calculation later in the paper,
these measurements are reported as Epochs 1 and 2 for each quasar,
with a code to indicate which survey supplied that epoch of data.
We report the date of each epoch as a Modified Julian Date (MJD).
For Mrk~231, we include on a second row the data of \cite{1978AJ.....83..566M}.

A word about NVSS flux densities is in order.
Z+06 used integrated flux densities to be conservative in selecting 
variable objects which were brighter in the FIRST survey.
GP07 used peak flux densities to obtain a more accurate measurement
of the variable radio flux density 
given the different resolutions of the two surveys.
However, for objects unresolved in NVSS, the integrated and peak flux densities
are the same (Eq. 42 of \citealt{con98}). All targets in Table 1 except Mrk 231
are unresolved by NVSS, and the NVSS peak flux densities
therein match the NVSS integrated flux densities reported by Z+06 for their
objects.  The flux densities reported by GP07 are the raw flux
densities, uncorrected for known biases, which ``should not normally be
used unless corrections are applied'' (\S 5.2.1 of \citealt{con98}).

The significance of the variability for each quasar
is calculated following Eq. (5) of GP07:
\begin{equation}
\sigma_{\rm var} = { F_F - F_N \over \sqrt{\sigma_F^2 + \sigma_N^2} }
\end{equation}
where the $F$ and $N$ subscripts denote the FIRST and NVSS surveys, 
respectively.
Six of the twenty-two BAL candidates reported in the literature
do not meet the $\sigma_{\rm var}>3$ criterion proposed by GP07 when
correct NVSS peak flux densities are used.  These objects are kept in the
table for completeness, but it should be kept in mind that their variability
is not formally statistically significant.
The variability of Mrk~231 between the NVSS and FIRST epochs was negative
and could be due to extended flux missed by FIRST.  For Mrk~231 we use
only the fluxes from \cite{1978AJ.....83..566M} in our subsequent analysis.

The other entries in Table 1 are discussed in the relevant sections 
below.

\section{Analysis} \label{analysis}
\subsection{Brightness Temperature \& Surface Brightness} \label{bright}

Following GP07, the brightness temperature in the quasar's cosmological 
reference frame, $T_{b,q}$, can be expressed
in terms of observable quantities on Earth, designated by the subscript $o$.
Consider a source in which the monochromatic intensity (surface brightness)
has changed by an amount $\Delta I_q(\nu_q)$ between epochs 1 and 2
separated by a time $\Delta t_q$.  In the Planck regime where $h \nu \ll k_BT$,
the change in brightness temperature associated with the change in intensity is
\begin{eqnarray}\label{DeltaTbq}
T_{b,2}-T_{b,1}\equiv\Delta T_{b,q}
= \frac{c^2 \Delta I_q(\nu_q) }{2 k_B \nu_q^2}
.
\end{eqnarray}
Intensity is equivalent to surface brightness.
To relate the change in the intrinsic surface brightness to the
observed flux densities, we assume a uniform source and use
the monochromatic version of Liouville's theorem (Eq. 30 of \citealt{gun78}):
\begin{eqnarray}\label{DeltaIq}
{ \Delta I_q(\nu_q) } =
{(1+z)^{3}} \Delta I_o(\nu_o) \equiv {(1+z)^{3}}
\left(\frac{F_{\nu,o,2}}{\Omega_2} - \frac{F_{\nu,o,1}}{\Omega_1}\right)
\end{eqnarray}
where $F_{\nu,o,1}$ and $F_{\nu,o,2}$ are the flux densities observed on Earth
at frequency $\nu_o=\nu_q/(1+z)$ at epochs 1 and 2,
separated by time interval $\Delta t_o=\Delta t_q(1+z)$.
The source subtends solid angle $\Omega_1$ at epoch 1 and $\Omega_2$ at epoch 2,
where the solid angle is the proper area of the
source on the sky divided by the angular diameter distance squared.

Equation \ref{DeltaIq} shows that flux density changes can occur without changes
in brightness temperature, if the emitting area changes in the same proportion
as the flux density.  That is, the flux density can change just because of 
a change in the emitting area of a source of constant surface brightness.

\subsection{Minimum Radii for a Source with $T_{b,q}<10^{12}$\,K} \label{min}

We can place a lower limit on the sizes of the emitting regions in these
quasars by finding the minimum sizes required for their brightness temperatures
to be $<10^{12}$\,K.
We assume a static source in this section,
and discuss the limitations of that assumption in later sections.

For a circular source of proper radius $r$ 
and observed flux density $F_{\nu,o}$ at a given epoch, 
Equations \ref{DeltaTbq} and \ref{DeltaIq} 
can be combined to write its brightness temperature at that epoch as
\begin{eqnarray}\label{Tbq1}
T_{b,q} = {c^2 \over 2k_B\nu_q^2}
\times { (1+z)^3 F_{\nu,o} \,d_A^2 \over \pi r^2 }
\end{eqnarray}
where $d_{A}$ is the angular diameter distance,
which is related to the luminosity distance $d_L$ as $d_A=d_L/(1+z)^2$.
GP07 adopt the expression $d_L=(1+z) cZ/H_0$ given by \citet{pen99}.
For $H_{0}$=70 km~s$^{-1}$~Mpc$^{-1}$,
$\Omega_{\Lambda}=0.7$ and $\Omega_{m}=0.3$, the quantity $Z$ is defined by
\begin{eqnarray}\label{Z}
Z \equiv 3.31-{3.65\over\left[z_1^{4}-0.203z_1^{3}+0.749z_1^2 +0.444z_1+0.205\right]^{1/8}}
\end{eqnarray}
where $z_1 \equiv 1+z$.

Combining all the above, $T_{b,q}$ can be written as
\begin{equation} \label{Tbq0}
T_{b,q} = \frac{c^2(1+z)^3}{2 k_B (1+z)^2 \nu_o^2}
\frac{ F_{\nu,o} (1+z)^2(cZ/H_0)^2 }{ \pi r^2 (1+z)^4 } \nonumber \\
\end{equation}
\begin{equation} \label{Tbq2}
T_{b,q} \pm \sigma_T
= \frac{
2.0\times 10^{12} (1+z) Z^2
} { (\nu_o/{\rm 1~GHz})^2 ([(1+z)r/c]/ {\rm 1~yr})^2 }
{ \left( F_{\nu,o} \pm \sigma_F \over \rm 1~mJy \right) }
{\rm ~K~}
\end{equation}
where 
$(1+z)r/c$ replaces $\Delta t_o/2$ in GP07 Eq. (3) and (4),
explaining the factor of four smaller normalization above.
%
The uncertainty on $T_{b,q}$ follows from assuming that the uncertainty on 
the flux density is dominant.
%

By setting $T_{b,q}=10^{12}$\,K in Eq. \ref{Tbq2} above, one can solve for
$r\equiv r^{\rm min}$ given the FIRST or NVSS $F_{\nu,o}$ measurement at 
$\nu_o=1.4$\,GHz.  The uncertainty on $r^{\rm min}$ is given by 
$\sigma_{r^{\rm min}}=r^{\rm min}\sigma_F/2F$.  Table 1 gives the resulting
values for both epochs 1 and 2, $r_1^{\rm min}$ and $r_2^{\rm min}$, in
proper light-years.
The sizes are reasonable for radio-emitting structures at the cores of
luminous AGN (e.g., \citealt{2004AJ....127..239G}).

\subsection{Expansion of an Emitting Source in the Plane of the Sky} \label{experp}

To explain these quasars using emission from plasma with $T_{b,q}<10^{12}$\,K
at both epochs requires reasonable minimum radii.
However, we must also determine whether the required expansion 
between the two epochs is reasonable.

We first consider whether sources expanding only in the plane of the sky
can explain the observations.
The minimum 
expansion velocity in the plane of the sky needed to explain the
observed variability by a source of constant brightness temperature
$T_{b,q}\leq 10^{12}$\,K is 
\begin{equation}\label{betaperpmin}
\beta_\perp^{\rm min} = ( r_2^{\rm min} - r_1^{\rm min} ) / c\Delta t_q .
\end{equation}
Table 1 gives values of $\beta_{\perp}^{\rm min}$ and associated uncertainties,
calculated from the uncertainties on the input $r^{\rm min}$.
There are four cases of contraction ($\beta_{\perp}^{\rm min}<0$)
--- one apparently superluminal --- which we consider in the next section, 
plus two cases for which $\beta_{\perp}^{\rm min}>1$.
In all other cases of expansion, the values of $\beta_{\perp}^{\rm min}$ 
are large but physically possible.

However, the values of $\beta_\perp^{\rm min}$ in Table 1 are for
uniform planar expansion in all directions in the plane of the sky.
A more likely scenario is a symmetric jet 
expanding along two directions in the plane of the sky
at $\beta\cong 1$ with opening angle $180^\circ f$,
instead of expanding at $\beta_\perp^{\rm min}$ as a filled disk;
i.e., opening angle $180^\circ$ in two opposite directions.
As the emitting area at the first and second epochs must be the same in both
cases, because such a jet covers only a fraction $f$ of a filled disk  
it must start from a radius a factor of $f^{-1/2}$ larger.
By assuming an expansion velocity of $c$ from such a radius, we find
$f^{\rm min}=(\beta_\perp^{\rm min})^2$.
Leaving out cases of contraction, $\beta_\perp^{\rm min}>1$, and
$\sigma_{\rm var}<3$, we find minimum opening angles of $4^\circ$ to 
$40^\circ$.  Such values are larger than those seen in quasars
with comparable radio luminosities, 
although they are similar to those found in weak radio galaxies \citep{bp84}.
Therefore, expansion close to the plane of the sky seems an unlikely 
explanation for the observed flux variability in these quasars,
even though it is physically possible.

\subsection{Spherical Expansion of an Emitting Source} \label{exp3D}

Spherical expansion of a radio-emitting source seems a more likely explanation.
Equation (\ref{Tbq2}) suggests that for a given source and $\nu_o$, an observed
flux density variability episode obeys 
$\Delta F_{\nu,o}(t) \propto T_{b,q}(t) [r(t)/r(0)]^2$,
where $r(t)$ is the apparent projected radius of the source on the sky 
at observed time $t$.  
Z+06 and GP07 assumed $r(t)=r(0)+ct$.
However, as pointed out by \citet{1966Natur.211..468R,1967MNRAS.135..345R}, 
an optically thick sphere expanding in all directions at $\beta=v/c$ increases
its projected area on the sky at a rate proportional to ($\gamma\beta$)$^2$, 
where $\gamma=(1-\beta^2)^{-1/2}$.
Such roughly spherical, expanding blobs of magnetized plasma could be launched
by a relativistic jet close to the plane of the sky in these objects.

Light travel time effects explain why a spherical emission source which is 
observed to expand between time $t=0$ to $t=t_o$ in our frame (time $t=0$ to
$t=t_q=t_o/(1+z)$ in the quasar frame) can appear to cover a larger area than 
$\pi(\beta ct_q)^2$, where we take $r(0)=0$ for simplicity in this example.
Consider the light from such a source
which reaches us at time $t_o$ in our frame: not all of it
was emitted at time $t_q$ in the quasar frame.
Some of that light was emitted after time $t_q$ in the quasar frame
from material which had travelled toward
us and decreased the light-travel time to reach us from it,
thus affording us a glimpse of a larger emitting source after a given observed
time interval than in the case of expansion solely in the plane of the sky.
The only requirement for that to happen is that
the emitting source must expand for a time $\gamma^2t_q$
in the quasar frame to be seen to expand to an area 
$\pi(\gamma\beta ct_q)^2$ over time $t_o$ in the observer frame.  
(To see this, note that
the part of the sphere which yields the largest apparent transverse motion
is that located at an angle $i=\arccos\beta$ to our line of sight.
During a time $t$ in the quasar frame, that material travels a distance
$\beta ct\cos i=\beta^2 ct$ toward Earth and light from the projected
center of the sphere travels a distance $ct$ toward Earth.  
Equating the difference in those distances 
to $c\Delta t_o/(1+z)=c\Delta t_q$ 
yields $t=\Delta t_q/(1-\beta^2)=\gamma^2 \Delta t_q$.)

Because $r(t)$$-$$r(0)$$\,\propto\,$$\gamma\beta t$ for spherical
expansion with $v$=$\beta c$, {\sl extremely large flux density increases
can be reproduced with $\beta$ sufficiently close to 1
without an intrinsic $T_{b,q}>10^{12}$\,K.}
In fact, the above equation is an underestimate for a source expanding with
$v\cong c$ because it does not account for all relativistic effects,
such as Doppler boosting in the part of the source expanding toward us.
Full accounting of such effects in a source whose outer edge expands with
velocity $\beta c$
shows that $\Delta F_{\nu,o}(t) \propto T_{b,q}(0) \gamma^{7/2} \beta^2 t^3$
(Eq. (8) of \citealt{1967MNRAS.135..345R}) in the first phase of the 
expansion.\footnote{The flux density grows more rapidly than $t^2$ even for 
nonrelativistic expansion because the source is assumed to be initially 
synchrotron self-absorbed at $\nu_q$, so that the intensity
of the emission at that frequency increases as the magnetic field
weakens in the expanding plasma; see Eq. (4) of \citealt{1967MNRAS.135..345R}.}
As the source starts to become optically thin to synchrotron self-absorption,
$\Delta F_{\nu,o}(t)$ peaks and then declines, initially roughly as 
$\gamma^{7/2}\beta^2t^{-1.5\pm0.5}$ \citep{1967MNRAS.135..345R}.
Therefore, {\sl extremely large flux density decreases
can also be reproduced with $\beta$ close to 1
without an intrinsic $T_{b,q}>10^{12}$\,K.}

To compute the required $\beta$ values, we compare to Z+06 and GP07,
who in effect assumed $v_0$=$c$ in Eq. 7 of \citet{1967MNRAS.135..345R}
to (incorrectly) estimate the brightness temperature
required for a source expanding at $\beta$$\cong$1 in the plane of the sky
to match the observed flux variations.
We denote those brightness temperatures with an asterisk
($\Delta T^*_{b,q}$) and give them in Table 1 in units of $10^{12}$\,K.
They are found by setting $r=c\Delta t_q$ in our Eq. \ref{Tbq2} and
using $ \Delta F_{\nu,o} = | F_{\nu,o,2} - F_{\nu,o,1} | $ and its uncertainty
in place of $ F_{\nu,o} $ and its uncertainty.
(The lower values of $\Delta T^*_{b,q}$ as compared to GP07 arise mainly
because the correct formula for $\Delta T^*_{b,q}$ given in Eq. 7
has a factor of 4 smaller normalization than given in GP07.)

The minimum $\beta$ required for spherical expansion to
explain the observed flux increases in these quasars depends on whether
the source responsible for the flux increase is pre-existing or new.
In the former case, the required $\beta$ is close to $\beta_\perp^{\rm min}$
except that it never exceeds unity.
In the latter case, the required $\beta$ can differ considerably from
$\beta_\perp^{\rm min}$.
We therefore compute the values $\beta_{r=0}^{\rm min}$ for the latter case;
these are the minimum $\beta$ required if spherical expansion from $r=0$ is to
explain the observations without requiring an intrinsic $T_{b,q}>10^{12}$\,K.
They are obtained by noting that in the first phase of the expansion, the
apparent $T_{b,q} \propto \Delta F_{\nu,o}(t)/[r(t)]^2 = \gamma^{3/2} t$.
Therefore, 
including relativistic effects means the intrinsic $T_{b,q}$ can be lower 
than the value $\Delta T^*_{b,q}$ calculated under the assumptions of Z+06 and
GP07 by a factor $X(\beta)$ of about $\gamma^{3/2}$, assuming a fixed fiducial
magnetic field strength in the sphere.\footnote{The exact factor $X(\beta)$ is
given by the ratio of $(F/\Omega)_{rel}$ to $(F/\Omega)_{\rm GP07}$,
where $(F/\Omega)_{rel}$ (surface brightness in the relativistic case)
is given by Eq. 6 of \citet{1967MNRAS.135..345R}
divided by $\pi(\gamma\beta ct)^2$,
and $(F/\Omega)_{\rm GP07}$ (surface brightness under the assumptions made by 
GP07) is given by Eq. 7 of \citet{1967MNRAS.135..345R}, with $v_0$ set to $c$
and a missing fraction of $\frac{10}{7}$ multiplied in, divided by $\pi(ct)^2$.}

Therefore, the observed flux variations can be matched by a source with
$T_{b,q}=10^{12}$\,K expanding spherically at $\beta$=$\beta_{r=0}^{\rm min}$,
where $\beta_{r=0}^{\rm min}$ is the solution to
$X(\beta_{r=0}^{\rm min})=\Delta T^*_{b,q}$.
Values of $\beta_{r=0}^{\rm min}$ are given in Table 1 for all cases where
$\Delta T^*_{b,q} \geq 10^{12}$\,K.

The values of $\beta_{r=0}^{\rm min}$ in Table 1 correspond to $\gamma<4.5$
in all but a few cases: $\gamma\simeq 5.5$ for J1346+3924, $\gamma\simeq 9$ for
J0828+3718, $\gamma\simeq 20$ for Mrk~231 and $\gamma\simeq 43$ for J0756+3714.
Values of up to $\gamma=50$ have been inferred for extragalactic jets 
\citep{mojaveVI}, but whether such values occur in cases of
spherical expansion of radio-emitting plasma is not clear.

Nonetheless, spherical expansion of an emitting source at
$\beta_{r=0}^{\rm min}$ or larger can explain all the radio variability
in these objects while still maintaining an intrinsic $T_{b,q}<10^{12}$\,K.
Values of $\beta$ above the minimum required would allow for shorter episodes
of variability to explain the same flux density changes,
which eases the requirement on how long the source must expand and remain
optically thick.

Larger values of $\beta$ could also maintain an intrinsic
$T_{b,q}<10^{11}$\,K.  For the three objects mentioned above,
the required $\beta$ values correspond to $\gamma\simeq 25$ for J1346+3924,
$\gamma\simeq 40$ for J0828+3718, $\gamma\simeq 94$ for Mrk 231
and $\gamma\simeq 200$ for J0756+3714.

Those quasars appear to require the most extreme parameters to explain their
variability without resorting to a jet directed close to our line of sight.
However,
underestimated flux errors could reduce the significance of the observed radio
variability of J1346+3924 ($\sigma_{\rm var}$=3.1) or J0756+3714 or Mrk~231
($\sigma_{\rm var}$=3.6) below a true 3$\sigma$ (99.7\% confidence) threshold, 
although that is extremely unlikely for J0828+3718 ($\sigma_{\rm var}$=10.4).

Furthermore, in the case of Mrk 231,
if there is a jet oriented near our line of sight then the lack of apparent
superluminal motion in the secondary VLBI component \citep{2009ApJ...706..851R}
means must that it be a near-stationary shock in that jet.
Otherwise, the jet would need to be oriented at $i<2^\circ$ of our line
of sight for the $\gamma$ values considered in \cite{2009ApJ...706..851R}.
Such small angles are extremely unlikely {\sl a priori}
($<$0.12\% chance if quasars are seen at all $i<60^\circ$).
They are also potentially ruled out in Mrk 231 by the detection of a rotating 
nuclear gas disk producing velocity gradients on the sky 
of $\pm$110\,\kms\ and $\pm$70\,\kms\ in 
\hi\ \citep{1998AJ....115..928C} and CO \citep{1996ApJ...457..678B} at radii 
of $<$85\,$h_{70}^{-1}$\,pc and $<$255\,$h_{70}^{-1}$\,pc, respectively.  Such
large observed gradients would imply unprecedentedly large intrinsic velocity
gradients for $i<2^\circ$, except in the case where the detected outer disk 
is misaligned with the innermost disk that sets the orientation of the jet
(see, e.g., \citealt{kgm05}).

We briefly consider additional effects that might affect
the likelihood of jets in these objects.
Varying the fiducial magnetic field in the radio-emitting region
cannot increase the limiting intrinsic $T_{b,q}$.
The model of \citet{1967MNRAS.135..345R} which we have used does
not include the additional surface brightness boost that would occur for a
relativistically expanding sphere that was also moving toward us at a
significant fraction of $c$.  However, the speed in our direction required to 
keep $T_{b,q}<10^{11}$\,K in J0756+3714 (e.g.) is 0.91$c$ (for expansion at
$\beta_{r=0}^{\rm min}$), which would require a relativistic jet oriented 
within $\leq 24^\circ$ of our line of sight.
Postulating two new radio-emitting regions, each responsible for half the
observed flux increase, can reduce $\Delta T^*_{b,q}$ by a factor of two
at most.  That could be significant for J0756+3714, Mrk 231 or J1346+3924,
given the large uncertainties on their inferred brightness temperatures,
but not for J0828+3718.
Overall, J0828+3718 is the best candidate for a true
polar BAL quasar.

\section{Conclusion} \label{conclusion}

Of the twenty-two candidate polar BAL quasars previously reported in the 
literature, only sixteen have statistically significant variability.
Those sixteen can be explained without requiring $T_{b,q}>10^{12}$\,K,
either by expansion of a pre-existing source at velocities of a few tenths of
lightspeed in most cases, or 
by a newly appeared spherical emitting source expanding at 
a lower limit $\beta_{r=0}^{\rm min}$ of $0.35c$ or greater.
To ensure $T_{b,q}<10^{11}$\,K would require larger velocities, but
only in one or two cases would the required velocity be unprecedentedly large.

Although a relativistic jet oriented close to our line of sight
is not required to explain the observed flux variability in any of these
quasars, we have not ruled out such a jet in any of them.
(However, note that a further observation of the candidate polar BAL quasar
SDSS J025625.56$-$011912.1 by \citet{2008MNRAS.388.1853M}
did not reveal the continued variability which is expected 
if a relativistic jet is oriented along our line of sight to that object.)
As two-epoch flux density variability is unable to unambiguously identify cases
of Doppler boosting, determining the relative incidence of relativistic jets
oriented along our line of sight in normal and BAL quasars will require
VLBI imaging to directly measure brightness temperatures.
In cases with $T_{b,q}>10^{12}$\,K, repeated imaging will be required to
determine whether such $T_{b,q}$ values are temporary (due to beaming from
relativistic expansion of plasma structures in intermittent flares)
or persistent (due to jet beaming)
and to constrain apparent transverse jet velocities in the latter case.
SDSS J082817.25+371853.7 is the best candidate for such imaging.

\section*{Acknowledgments}
PBH and LSC are funded by NSERC.  PBH also thanks the Institute of Astronomy at
the University of Cambridge for funding and hospitality during his sabbatical.
We thank B. Punsly for bringing Reynolds et al. (2010) to our attention.


\begin{figure*}
\makebox[\textwidth]{
\includegraphics[angle=180, width=1.060\textwidth]{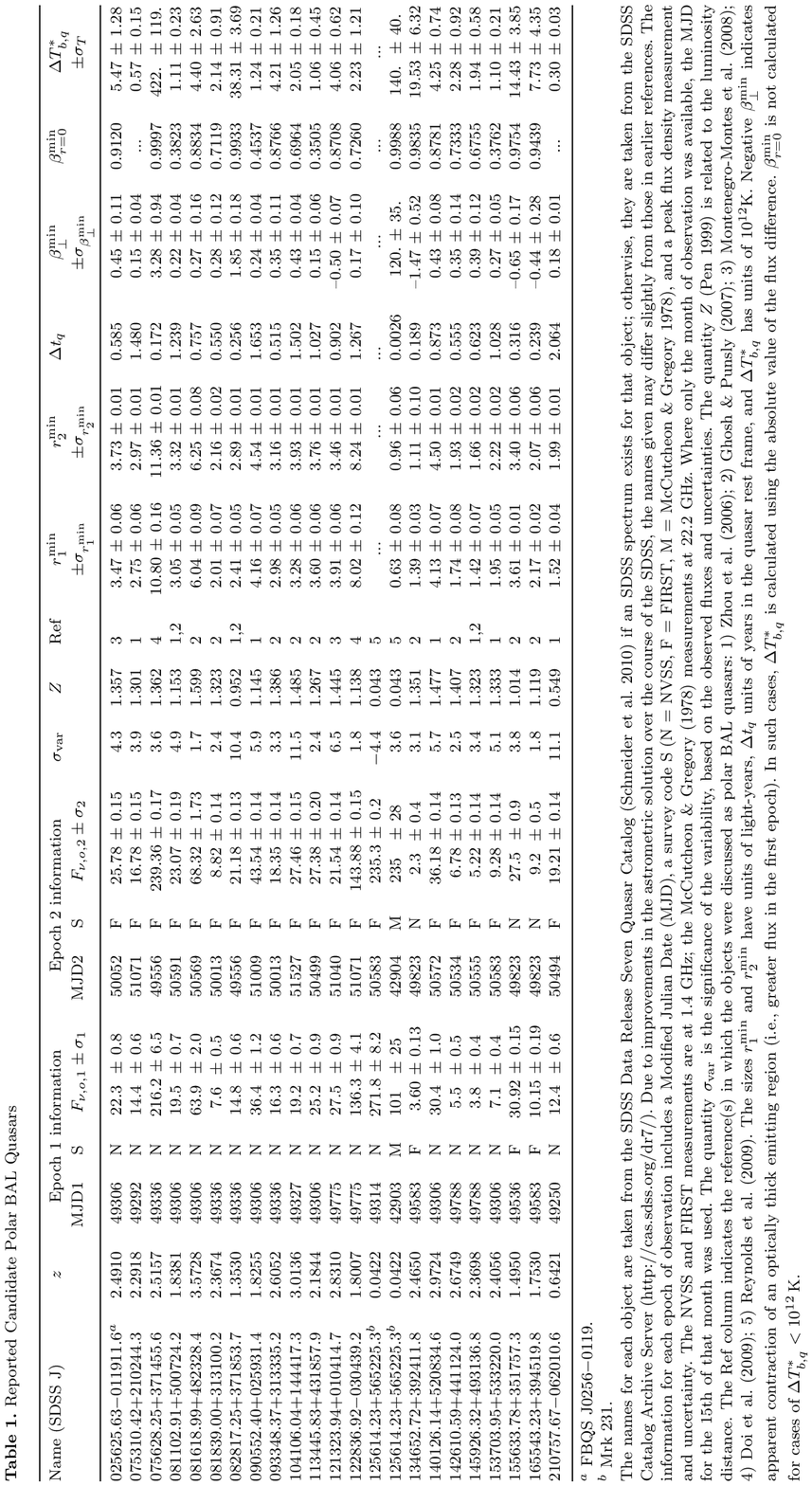}}
  \caption{}
  \label{landfig}
\end{figure*}



\label{lastpage}

\end{document}